\documentstyle[prd,preprint,tighten,aps]{revtex}
\begin{document}
\draft
\preprint{
\begin{tabular}{r}
DFTT 3/96
\\
hep-ph/9601389
\end{tabular}
}
\title{On the impossibility to distinguish
MSW transitions from vacuum oscillations
for neutrino maximal mixing}
\author{
S.M. Bilenky$^{\mathrm{a}}$
and
C. Giunti$^{\mathrm{b}}$
}
\address{
\begin{tabular}{c}
$^{\mathrm{a}}$Joint Institute for Nuclear Research,
Dubna, Russia,
and
\\
SISSA-ISAS, Trieste, Italy.
\\
$^{\mathrm{b}}$INFN,
Sezione di Torino and Dipartimento di Fisica Teorica,
Universit\`a di Torino,
\\
Via P. Giuria 1, 10125 Torino, Italy.
\end{tabular}
}
\date{February 1, 1996}
\maketitle
\begin{abstract}
We show in a simple and general way
that MSW resonant transitions
for solar neutrinos are not observable
in the case of maximal mixing among
any number of neutrino generations.
\end{abstract}


\narrowtext

There has been recently
much interest in the case of maximal mixing
among neutrino generations.
The existing data of the
solar neutrino experiments
and of the atmospheric neutrino experiments,
in which indications in favor of neutrino
mixing were found,
have been analyzed
in Refs.\cite{GKK95,HPS95}
under the assumption
of maximal mixing
among three generations of neutrinos.

In a recent preprint
\cite{HPS96}
it has been noticed that,
in the case of maximal mixing among two or three
neutrino generations,
even if the conditions for MSW
\cite{MSW}
resonant transitions in the sun are satisfied,
their effects are not observable.
This conclusion was based on a numerical
solution of the evolution equation.
We present here a simple and general
derivation of this result
in the case of mixing among
any number $N$ of neutrino generations.

Let us consider the case of an arbitrary number $N$
of massive neutrinos.
If there is neutrino mixing
(see Ref.\cite{Bilenky})
the neutrino fields
$\nu_{{\alpha}L}$
are given by a superposition
of (Dirac or Majorana)
fields
$\nu_{aL}$
with mass $m_a$ as
\begin{equation}
\nu_{\alpha L}
=
\sum_{k=1}^{N}
U_{\alpha a}
\nu_{aL}
\;.
\label{100}
\end{equation}
Here $U$ is a unitary mixing matrix.

Let us now consider solar $\nu_e$'s
that are produced in
the thermonuclear reactions
occurring in the core of the sun.
As a result of neutrino propagation
from the core of the sun to its surface,
which can include possible
MSW resonant transitions,
the neutrino state emerging
from the surface of the sun
is a superposition of neutrino states
which can be written as
\begin{equation}
\left| \nu_{\odot} \right\rangle
=
\sum_{\alpha}
A_{e\alpha}(E)
\left| \nu_{\alpha} \right\rangle
\;,
\label{11}
\end{equation}
with
\begin{equation}
\sum_{\alpha}
\left| A_{e\alpha}(E) \right|^2
=
1
\;.
\label{12}
\end{equation}
The amplitudes
$ A_{e\alpha}(E) $
are determined by
the evolution equation of neutrinos
in the sun.
The values of these amplitudes will be not
important in this note,
we will use only the normalization condition (\ref{12}).
Let us stress only that if MSW matter effects are important
the amplitudes
$ A_{e\alpha} $
depend on the neutrino energy $E$.

The neutrino state on the earth
is given by
\begin{equation}
\left| \nu_{\oplus} \right\rangle
=
\sum_{\alpha,a,\rho}
A_{e\alpha}
U_{{\alpha}a}^{*}
\mbox{e}^{-i E_a T}
U_{{\rho}a}
\left| \nu_{\rho} \right\rangle
\;,
\label{111}
\end{equation}
where $T=R$ and
$R$ is the distance between the
surface of the sun and the earth.

The probability of transitions of
solar $\nu_e$'s into any state
$\nu_{\rho}$
at the distance $R$
is given by
\begin{equation}
P_{\nu_e\to\nu_\rho}
=
\sum_{\alpha,\beta,a,b}
A_{e\alpha}
\,
A_{e\beta}^{*}
\,
U_{{\alpha}a}^{*}
\,
U_{{\rho}a}
\,
U_{{\beta}b}
\,
U_{{\rho}b}^{*}
\,
\exp
\left(
- i
\,
{\displaystyle
\Delta m^2_{ab}
\over\displaystyle
2 E
}
\,
R
\right)
\;,
\label{13}
\end{equation}
where
$ \Delta m^2_{ab} \equiv m^2_a - m^2_b $.

We will assume that
\begin{equation}
\left| \Delta m^2_{ab} \right|
\gg
10^{-10} \, \mbox{eV}^2
\label{101}
\end{equation}
for all values of $a$ and $b$ ($ a \not= b $).
This inequality is satisfied
if there are MSW resonances in the sun.
From Eq.(\ref{13}),
for the measurable
averaged probability we have
\begin{equation}
\left\langle P_{\nu_e\to\nu_\rho} \right\rangle
=
\sum_{a}
\left| U_{{\rho}a} \right|^2
\left|
\sum_{\alpha}
A_{e\alpha}
\,
U_{{\alpha}a}^{*}
\right|^2
\;.
\label{14}
\end{equation}

In the case of maximal mixing
$ \left| U_{{\alpha}a} \right|^2 = 1/N $
(see Refs.\cite{GKK95,HPS95}).
Using the unitarity of the mixing matrix
and the normalization condition (\ref{12})
we obtain
\begin{equation}
\left\langle P_{\nu_e\to\nu_\rho} \right\rangle
=
{ 1 \over N }
\sum_{a}
\left|
\sum_{\alpha}
A_{e\alpha}
\,
U_{{\alpha}a}^{*}
\right|^2
=
{ 1 \over N }
\sum_{\alpha}
\left| A_{e\alpha} \right|^2
=
{ 1 \over N }
\;.
\label{15}
\end{equation}
Thus,
in the case of maximal mixing,
if the condition (\ref{101})
is satisfied,
the probability of transitions of
solar $\nu_e$'s into any
state $\nu_\rho$,
active or sterile,
do not depend on $\nu_\rho$ and is equal to $1/N$.
This probability does not depend
on the values
of the amplitudes
$ A_{e\alpha} $,
which means that the MSW effect is not observable
in the case of maximal mixing.
In Ref.\cite{HPS96}
this result was obtained numerically for
the cases $N=2$ and $N=3$.

\acknowledgments

S.B. would like to acknowledge
the kind hospitality of the
Department of Theoretical Physics
of the University of Torino,
where this work has been done.

\end{document}